# Power Converters and Power Quality

*K. Kahle*
CERN, Geneva, Switzerland

**Abstract**
This paper discusses the subject of power quality for power converters. The first part gives an overview of most of the common disturbances and power quality issues in electrical networks for particle accelerators, and explains their consequences for accelerator operation. The propagation of asymmetrical network disturbances into a network is analysed. Quantitative parameters for network disturbances in a typical network are presented, and immunity levels for users' electrical equipment are proposed. The second part of this paper discusses the technologies and strategies used in particle accelerator networks for power quality improvement. Particular focus is given to networks supplying loads with cycling active and reactive power.

## 1 Introduction

Network disturbances and power quality are issues of increasing importance, as the share of sensitive electronic equipment is steadily increasing in modern power systems. Due to their complex nature and the highest requirements for precision, power converters for particle accelerators are extremely sensitive to network disturbances. A large percentage of all unwanted accelerator stops, so-called major events, is related to such disturbances. Several cases of particle beam instabilities that were caused by power quality issues have been observed. This paper discusses the principal aspects of power quality, with a particular focus on power converters for particle accelerators.

The main objectives of this paper are to raise awareness amongst power converter specialists about this subject, and to discuss the technologies and main engineering principles that can be applied to reduce the impact of power quality issues on particle accelerator operation.

## 2 Classification of disturbances

The most critical disturbances are discussed below [1-9].

### 2.1 Voltage dips

Voltage dips are two-dimensional phenomena, characterized by their amplitude and duration. They are typically caused by short-circuits in the electrical network upstream or downstream of the observation point. Voltage dips have a typical duration of 50–200 ms (in some cases up to 1000 ms), and an amplitude typically of −5% to −100%.

Voltage dips are the principal cause of power quality issues related to particle accelerators, and for this reason it is essential to define the immunity levels of the electrical equipment in order to 'survive' most of these disturbances without tripping the particle accelerator.

## 2.2 Short supply interruptions

Short supply interruptions can be seen as voltage dips to zero. They are typically caused by short-circuits in overhead lines, which are cleared when the protection systems reclose. Short supply interruptions can last up to 180 s. For short supply interruptions, the concept of immunity of equipment is not applicable, simply because most equipment will not be able to function without its energy supply even for short periods of time.

## 2.3 Flicker

Flicker is defined as the impression of unsteadiness of visual sensations induced by a light stimulus whose luminance fluctuates with time. A typical example is the visible change in the brightness of a lamp due to rapid and repetitive fluctuations of the network voltage. These voltage fluctuations are typically caused by variations in load current passing through the source impedance of the electrical network upstream.

The repetitive active and reactive power cycles of the main power converters of a particle accelerator are a typical cause of flicker. The voltage drop caused by the load current flowing through the impedances of the upstream network has a major influence on the voltage amplitude and phase angle at a particular busbar. Depending on the network inductances and resistances upstream, as well as the amplitude of active and reactive power of the load, the voltage drop is calculated as follows:

$$\Delta U = RI \cos\phi + XI \sin\phi, \quad (1)$$

$$\frac{\Delta U}{U} = \frac{R.\Delta P + X.\Delta Q}{U^2}. \quad (2)$$

Equation (2) shows that any variation of load active and reactive power, as well as any change in network inductance or resistance, will result in a change in voltage drop and hence a variation of operating voltage at a given busbar. The most prominent example is the cycling power of thyristor converters for synchrotron accelerators.

The voltage fluctuations are defined as changes of r.m.s. voltage evaluated as a single value for each successive half-period of the network voltage. The short-term flicker indicator $P_{st}$ specifies the severity of the flicker evaluated over a short period (e.g. 10 min); and $P_{st} = 1$ is the conventional threshold of irritability. The long-term flicker indicator $P_{lt}$ defines the flicker severity evaluated over a long period (typically 2 h) using the cubic average of successive $P_{st}$ values. Typically, the admissible long-term flicker factor is 60% of its short-term equivalent. Figure 1 shows the limits for a short-term flicker factor of 1.

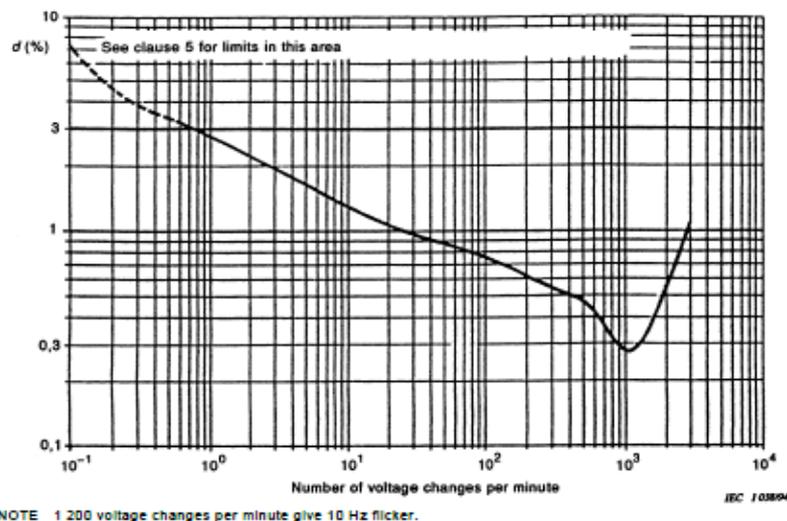

NOTE 1 200 voltage changes per minute give 10 Hz flicker.

**Fig. 1**: Flicker limits for short-term flicker factor $P_{st} = 1$ (source: IEC 61000-3-3, Fig. 4)

These flicker limits are based on the empiric definitions of the human eye's sensitivity to luminance fluctuations. At the point of connection with the external network, the flicker level represents a contractual constraint and needs to be strictly respected.

The irritating effects of flicker can be reduced by applying a combination of the following measures:

- strictly separate general services and power converter loads;
- reduce the network impedance (increase short-circuit power) at the load busbar;
- reduce the amplitude of the (active and) reactive power variation;
- install technologies for flicker mitigation.

## 2.4 Fast transient overvoltages

Fast transient overvoltages (surges) are typically caused by lightning strikes into overhead lines, hence having a much lower probability in buried cable networks. They can also be caused by network switching operations, such as switching off large or small inductive loads, as well as short-circuits. The amplitude of fast transient overvoltages can easily reach multiples of the nominal system voltage, and hence they need to be taken into consideration for insulation coordination of the electrical network. For the power equipment itself, an appropriate choice of component voltage rating combined with simple mains input filters are suitable measures to obtain immunity against fast transient overvoltages.

## 2.5 Temporary power frequency variations

Temporary power frequency variations are less critical for most electrical equipment, and there are no specific countermeasures required.

The fundamental frequency of an electrical network is an expression of balance between power generation and consumption:

$$dE_k/dt = P_g - P_c ,\qquad(3)$$

where $E_k$ is the kinetic energy of all rotating machines ($= \frac{1}{2} J \omega^2$); $P_g$ is total generated power, $P_c$ – total consumed power; $J$ is the torque of the rotating machines; and $\omega$ is angular speed (rad/s).

The frequency of the European 400 kV network of the Union for the Co-ordination of Transmission of Electricity (UCTE) network is closely controlled, with typical fluctuations of only ±0.1 Hz around the nominal value of 50 Hz. The overall network power frequency characteristic of the 400 kV UCTE network is about 27 000 MW/Hz. As a simplified example (static point of view), an increase of power consumption by 270 MW would cause a drop in frequency of 10 mHz, if the power generation was not simultaneously increased.

Due to its three-level frequency control system, the large scale of generation systems, and the synchronous interconnection with other HV networks on a continental scale, the UCTE network has ample reserves to cope with load power variations under normal operating conditions.

However, in rare cases of breakdown of major interconnections and/or large groups of generation systems, the stability and continued operation of the entire network might be at risk. In such situations, the network operators could allow larger frequency variations in order to prevent runaways that would result in large-scale blackouts. In such exceptional situations the temporary frequency variations could be in the order of several hertz, being the result of larger imbalances between power generation and consumption.

## 2.6 Harmonics

Harmonics are conducted low-frequency disturbances of integer multiples of the system frequency. According to IEC standards [3–9], the total harmonic distortion is calculated taking into account all harmonics up to the 50$^{th}$ harmonic order.

The principal causes of harmonics in electrical networks are non-linear loads such as large thyristor power converters or diode rectifiers. During transformer energization, the transformer magnetizing currents (inrush) also generate significant temporary harmonic distortion.

The generated harmonic spectrum of thyristor power converters on the AC input side is as follows:

$$\text{Order of harmonics} = n \times p \pm 1 , \tag{4}$$

where $n = 1, 2, 3\ldots$ and $p$ = 6- or 12-pulse converters.

Harmonic distortion typically changes only slowly over time for stable loads; however, for fast cycling loads such as pulsating particle accelerators they change rapidly during each power cycle. For rapidly changing harmonics the conventional definitions according to the IEC are not meaningful, and smaller time windows (e.g. 100 ms) for the calculation of the harmonic distortion need to be defined.

These harmonic currents create harmonic voltage drops when passing through the upstream network impedances, which are then superimposed upon the fundamental frequency network voltage. These are the most common techniques to reduce the harmonic levels in a network:

– strict separation of general services and power converter loads;
– reduction of source impedance (increase short-circuit power) at the load busbar;
– installation of passive or active harmonic filters (see Section 6.1.1).

## 2.7 Temporary overvoltage

A temporary overvoltage has a typical duration of 10–500 ms, and can occur during a single-phase earth fault in the healthy phase(s) of a medium voltage network with non-effectively grounded neutral. The voltage increase is typically between +30% and +70%. Temporary overvoltages also occur during the energization of large capacitor banks, where the capacitive inrush current passing through the inductances of the upstream network causes a temporary voltage increase. These overvoltages have a much shorter duration of 10–20 ms, and typical amplitude of up to +50%.

## 2.8 Voltage imbalance

In the event of a voltage imbalance, the fundamental frequency voltage amplitude or angle between phases are not equal for all three phases. The imbalance is usually expressed as the ratio of negative to positive sequence components, using the following approximation:

$$\text{Voltage imbalance} = \sqrt{6 \frac{U_{12}^2 + U_{23}^2 + U_{31}^2}{U_{12} + U_{23} + U_{31}} - 2} , \tag{5}$$

where $U_{12}$, $U_{23}$ and $U_{31}$ are the fundamental frequency line–line voltages.

## 3 Power quality statistics

It is important to have a good understanding of the power quality of the electrical network. In particular, the duration and amplitude of voltage disturbances play a critical role in accelerator operating reliability. Statistics for the CERN electrical networks are presented below as an example [10].

### 3.1 Voltage dips

Table 1 shows the number of voltage dips recorded during the reference period (mid June–mid November) at CERN.

**Table 1:** Number of voltage dips recorded at different voltage levels (period mid June–mid November)

| Network | Substation | Machine | Number of voltage dips |
|---------|------------|---------|------------------------|
| 400 kV  | BE         | All of CERN | 37 |
| 18 kV   | EMD1/BE    | SPS     | 104 |
| 0.4 kV  | ERD1/8R    | LHC     | 184 |

Network disturbances at the 400 kV level are usually seen also in the 18 kV and 0.4 kV networks, while disturbances generated at the 0.4 kV level do not propagate upwards into the HV networks.

Network disturbances at CERN that cause one of the accelerators to stop are called 'major events', and are marked in red in Figs. 2–4. Although there are only limited statistical data available, it seems that the duration of network disturbances is not the critical parameter for accelerator trip; rather, it is the voltage amplitude. The data also indicate that most disturbances causing major events originate from the 400 kV network. Figures 2–4 show the statistics for network disturbances over a reference period (mid June–mid November):

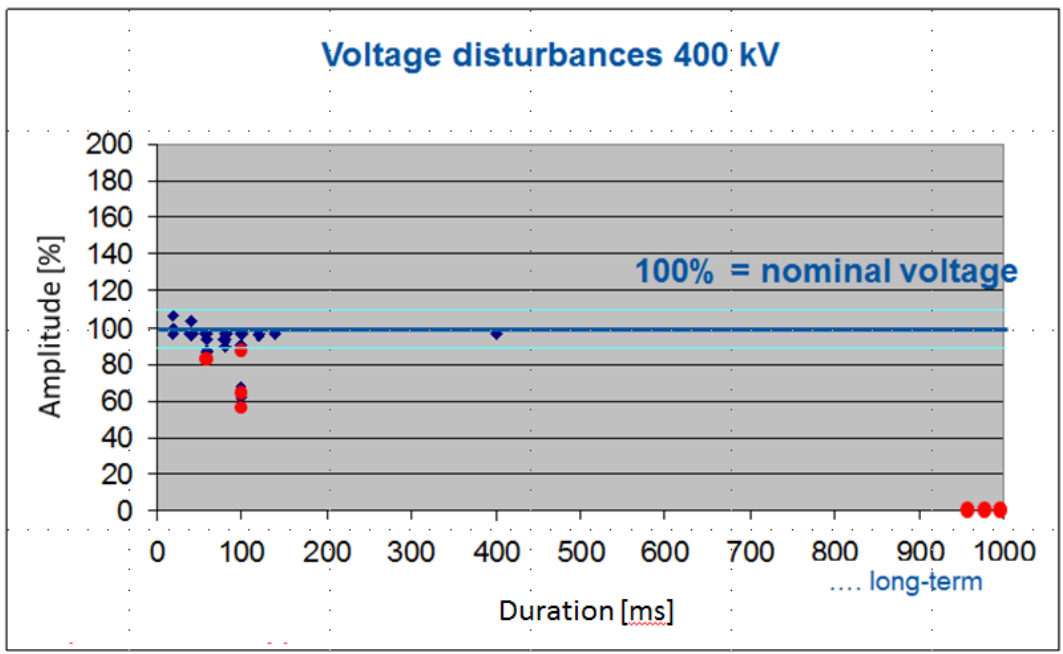

**Fig. 2:** Recorded disturbances in the 400 kV network. The events marked in red are major events, leading to a stoppage of an accelerator.

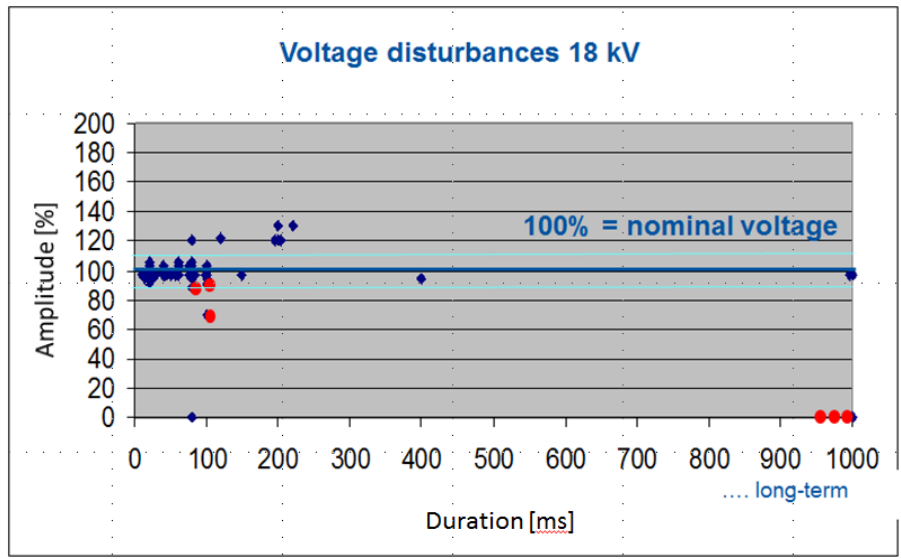

**Fig. 3:** Recorded disturbances in one of the 18 kV networks. The events marked in red are major events, leading to a stoppage of an accelerator.

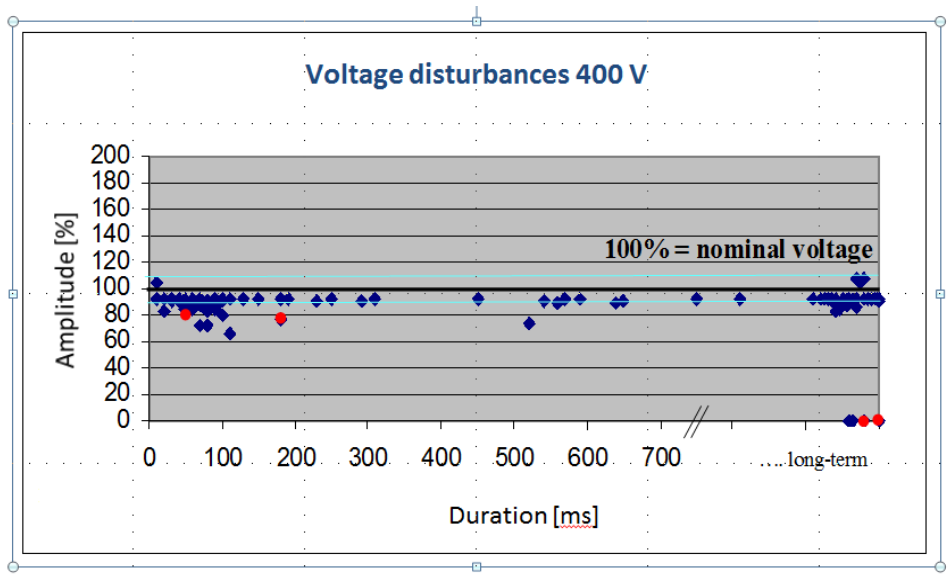

**Fig. 4:** Recorded disturbances in one of the 400 V networks. The events marked in red are major events, leading to a stoppage of an accelerator.

### 3.2  Propagation of external 400 kV voltage disturbances into the downstream voltage levels

Symmetrical 400 kV network disturbances propagate symmetrically into the subsequent (lower) voltage levels, and the relative amplitude of the disturbance remains the same for all voltage levels.

This article explains the propagation of asymmetrical 400 kV network disturbances into the downstream voltage levels, as found with computer simulations. All disturbances originating from the 400 kV network are transmitted to the lower voltage levels according to the vector groups of the transformers. Figure 5 graphically represents the vector relationships between the different voltage levels.

Based on the vector diagrams in Fig. 5, the asymmetrical disturbances propagate into the lower voltage levels (Tables 2, 3, 4 and 5). During this propagation, the faulty phase is partially recovering, while the healthy phases experience a decrease in voltage amplitude. Due to this effect, the symmetry of the three phases is improved even during asymmetrical conditions. For example, a 50% voltage dip in phase R of the 400 kV network results in a 66% dip in phase T of the LV network, and 94% dips in phases R and S. The causes of this effect are the vector groups of the transformers and the reduction of zero sequence voltage components during the transformation 400/66/18 kV, and thus the symmetrization of the three phase voltages. During the transformation 18/0.4 kV there is no reduction of zero sequence voltage components (0.4 kV starpoint directly grounded) and thus no symmetrization effect.

**Table 2:** Propagation of single-phase voltage dip in 400 kV network, −50% in phase R

|  |  | R | S | T | R-S | S-T | R-T |
|---|---|---|---|---|---|---|---|
| 400 kV |  | 50% | 100% | 100% | 75% | 100% | 75% |
| 66 kV |  | 58% | 97% | 96% | 78% | 100% | 77% |
| 18 kV |  | 77% | 100% | 77% | 95% | 96% | 65% |
| 18/0.4 kV | 0.4 kV | 94% | 94% | 66% | 100% | 77% | 77% |
| 18/3.3/0.4 kV | 3.3 kV | 94% | 94% | 66% | 100% | 75% | 78% |
|  | 0.4 kV | 94% | 94% | 66% | 100% | 76% | 76% |

**Table 3:** Principle of propagation of single-phase voltage dip from 400 kV to 0.4 kV

| Voltage level | Faulty phase |
|---|---|
| 400 kV | R |
| 66 kV | R |
| 18 kV | R-T |
| 3.3 kV | T |
| 0.4 kV | T |

**Table 4:** Propagation of double-phase dip in 400 kV network, −50% in phases R and T

|  |  | R | S | T | R-S | S-T | R-T |
|---|---|---|---|---|---|---|---|
| 400 kV |  | 50% | 97% | 50% | 76% | 76% | 50% |
| 66 kV |  | 57% | 87% | 58% | 76% | 76% | 50% |
| 18 kV |  | 76% | 76% | 50% | 83% | 60% | 60% |
| 18/0.4 kV | 0.4 kV | 83% | 60% | 60% | 77% | 50% | 77% |
| 18/3.3/0.4 kV | 3.3 kV | 84% | 66% | 64% | 77% | 53% | 77% |
|  | 0.4 kV | 83% | 65% | 65% | 78% | 54% | 78% |

**Table 5:** Principle of propagation of double-phase voltage dip from 400 kV to 0.4 kV

| Voltage level | Healthy phase |
|---|---|
| 400 kV | S |
| 66 kV | S |
| 18 kV | R-S |
| 3.3 kV | R |
| 0.4 kV | R |

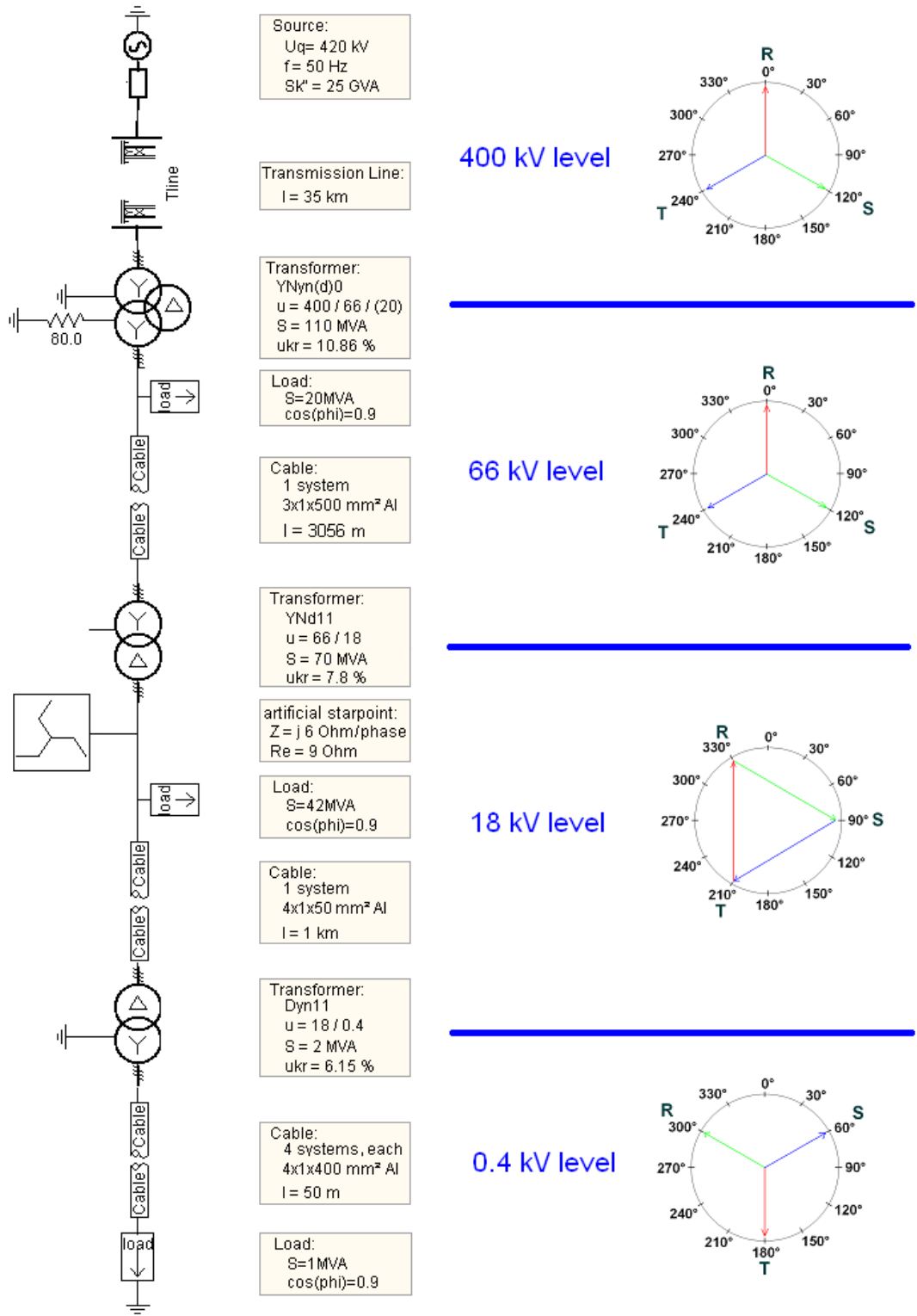

**Fig. 5:** Simplified CERN network and vector relations between voltage levels

In the case of a single-phase interruption of the 400 kV network, the voltage in the faulty phase remains close to 100% due to the 400/66 kV vector group YNyn with delta compensation winding. In this case, the compensation winding provides the magnetizing flux for the missing phase such that the voltage remains close to 100% (Table 6).

**Table 6:** Propagation of single-phase interruption 400 kV (after opening of breaker, phase R)

|        | R     | S    | T    | R-S  | S-T  | R-T  |
|--------|-------|------|------|------|------|------|
| 400 kV | (97%) | 100% | 100% | 100% | 100% | 100% |
| 66 kV  | 91%   | 102% | 95%  | 100% | 100% | 89%  |
| 18 kV  | 99%   | 100% | 89%  | 103% | 93%  | 92%  |
| 0.4 kV | 103%  | 94%  | 91%  | 100% | 89%  | 98%  |

Single-phase short-circuits on the 400 kV overhead line from Genissiat to BoisTollot caused, for example, by a flash-over due to lightning strike, result in a single-phase voltage dip in the faulty phase and an increase of voltage amplitude in the two healthy phases. The voltage rise of the healthy phases is related to the distance, and increases if the short-circuit happens closer to CERN. For a single-phase short-circuit directly at BoisTollot, the impedance ratio $Z_0/Z_1$ is the highest; and the worst-case voltage increase in the healthy phases could reach up to +17% (voltage phase–ground) when CERN is only supplied from Genissiat 400 kV. In the case of parallel infeed from Genissiat and Chamoson, the voltage rise in the healthy phases is +11% during a single-phase short-circuit. Again, the propagation into CERN follows the vector relationships as described above.

### 3.3 Harmonic voltage distortion

Repeated measurements of harmonic distortion were done in different 0.4 kV substations in CERN's electrical distribution network, showing that in all network supplied substations the harmonic distortion remained below 5%, corresponding to IEC 61000-2-2 (class 1). Generally, harmonics are lowest in switchboards for general services (EBDxx/xx). Higher harmonic voltage distortions are present in switchboards supplying rectifier loads (ERDxx/xx). At one UPS-supplied EODxx/xx switchboard, a harmonic distortion of 8% was measured. Figure 6 shows the total harmonic voltage distortion (THD) measured at different LV substations at CERN during accelerator operation.

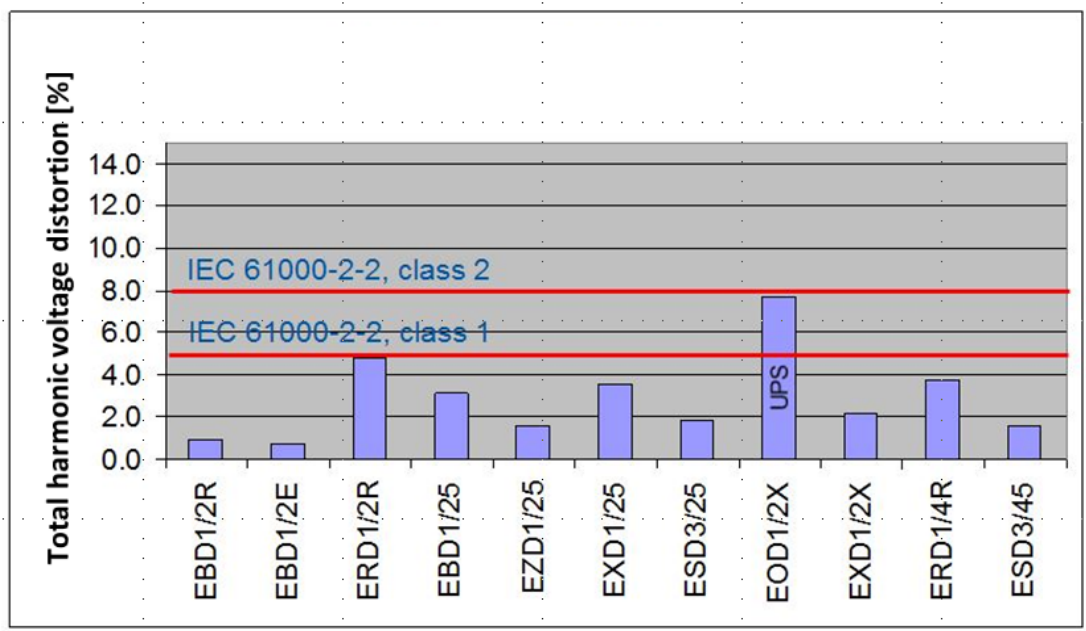

**Fig. 6:** Measured harmonic distortion at different LV substations

## 4 Electromagnetic environment classes

The electromagnetic environment classes are defined by IEC 61000-2-4:

– class 1: protected supplies for compatibility levels lower than those on public networks… for very sensitive equipment;

- class 2: environments of industrial and other non-public power supplies… and generally identical to public networks;
- class 3: industrial environments, in particular when a major part of the load is power converters and/or the load varies rapidly (e.g. particle accelerators!).

Table 7 quantifies what these three classes actually mean, and also compares them with the LHC Engineering Specification for LV networks [11] and the example of a static Var compensator (SVC) from the SPS machine network.

Table 7: Parameters of the electromagnetic environment classes (IEC 61000-2-4)

|  | Class 1 | Class 2 | Class 3 | LHC Engineering Specification [11] | Example: SVC for SPS (18 kV) |
|---|---|---|---|---|---|
| Voltage tolerances | ±8% | ±10% | −15%/+10% | Typical ±5%, maximum ±10% | ±0.75% (transient) |
| THD (400 V) | 5% (short-term 7.5%) | 8% | 10% (short-term 15%) | Typically 2%, maximum 5% | 0.75% (transient) |
| Frequency tolerances | ±1 Hz | ±1 Hz | ±1 Hz | ±0.5 Hz | ±0.5 Hz |

Power converters for particle accelerators represent the roughest type of load, comparable to heavy industry such as large arc furnaces, rolling mills, etc. (class 3). However, to operate them correctly and with the required precision, power converters for particle accelerators require power quality levels sometimes better than the most sensitive equipment (class 1)!

The most common technologies for power quality improvement, used for particle accelerators, are explained in the second part of this publication.

## 5  Definition of immunity levels for equipment (LHC Engineering Specification)

Based on the statistics for past network disturbances, combined with considerations of technical feasibility, the minimum immunity levels for electrical equipment were defined before the beginning of LHC construction [11], covering the main parameters of CERN's low voltage systems, their variations, and power quality issues. Users' equipment should be designed to correctly function within the limits detailed in Table 8.

Table 8: User equipment specifications

| | |
|---|---|
| Nominal voltage | 400 V/230 V |
| Maximum voltage variations | ±10% |
| Typical voltage variations | ±5% |
| Transients (spikes) | 1200 V for 0.2 ms |
| Voltage swells | +50% of $U_n$, 10 ms |
| Voltage dips | −50% of $U_n$, 100 ms |
| Total harmonic voltage distortion (THD) | 5% |

These voltage limits are visualized in Fig. 7:

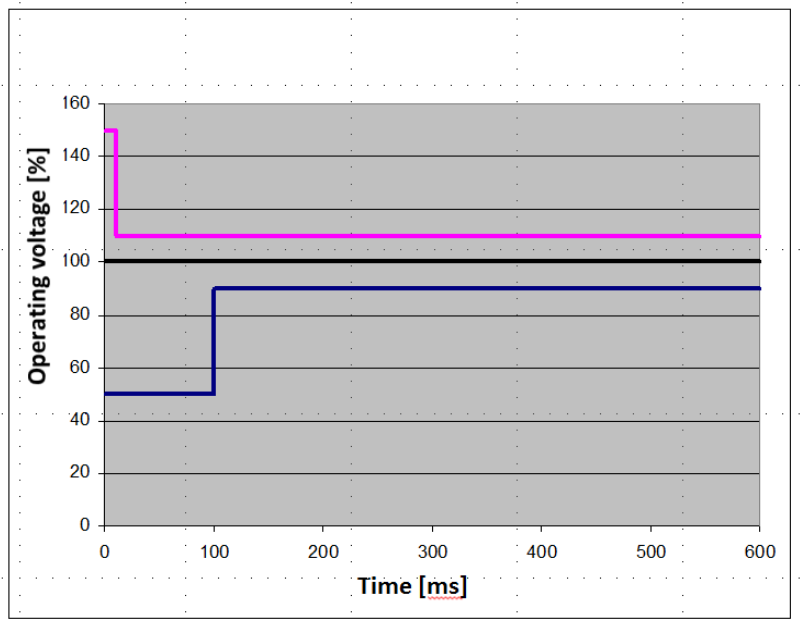

**Fig. 7**: Immunity levels for voltage disturbances as per the engineering specification for LV networks

## 6 Technologies for power quality improvement

This article discusses measures and technologies used in accelerator networks for power quality improvement. Particular focus is given to the improvement of power quality in networks supplying cycling loads.

### 6.1 Systems without integrated energy storage

#### *6.1.1 Static Var compensators*

Static Var compensators (SVCs) are part of a larger classification of equipment called flexible AC transmission systems (FACTS). An SVC typically consists of a combination of harmonic filters for the generation of (capacitive) reactive power and for harmonic filtering, as well as thyristor controlled reactors (TCR) to allow for the control of variable reactive power output. Figure 8 shows a very simplified topology for an SVC, consisting of a fixed capacitance (e.g. harmonic filters) and a TCR [12, 13].

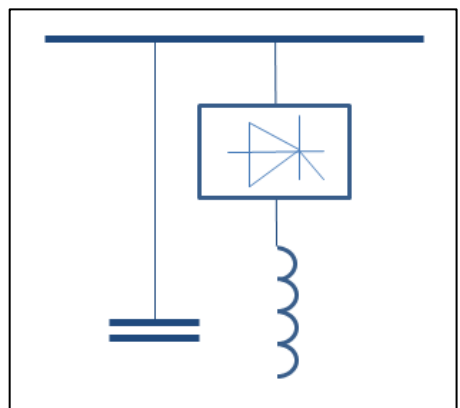

**Fig. 8:** Very simplified topology of an SVC

Optimum performance is achieved by connecting the SVC to the same MV substation as the cycling power converter load, typically directly downstream of the feeding HV/MV transformer. In most cases, the MVar ratings of the harmonic filters and of the TCR are identical; however, for certain applications a slightly increased TCR might be an interesting option.

The current in capacitor banks (or harmonic filters) cannot be continuously controlled by thyristors; capacitors can only be switched on or off, e.g. by means of thyristors or mechanical switches. The step-wise switching of individual capacitor banks (or harmonic filters) would generate transient overvoltages and other disturbances in the electrical network, which could potentially disturb the operation of the particle beam. Instead, and to follow the cycling reactive power of the load, the current in the thyristor-controlled reactors is adapted continuously by variation of the firing angle of the thyristors. Therefore, the capacitor banks (or harmonic filters) permanently generate a constant amount of reactive power, while the TCR follows the load pulse. As shown in Fig. 9, the reactive power of the TCR is controlled in the opposite way with respect to the reactive power cycles of the load: during the flat top of the load cycle the TCR current is low, and during low load the TCR current is at maximum. Figure 19 shows the qualitative functions of reactive power, while Fig. 13 represents measurements of a real system.

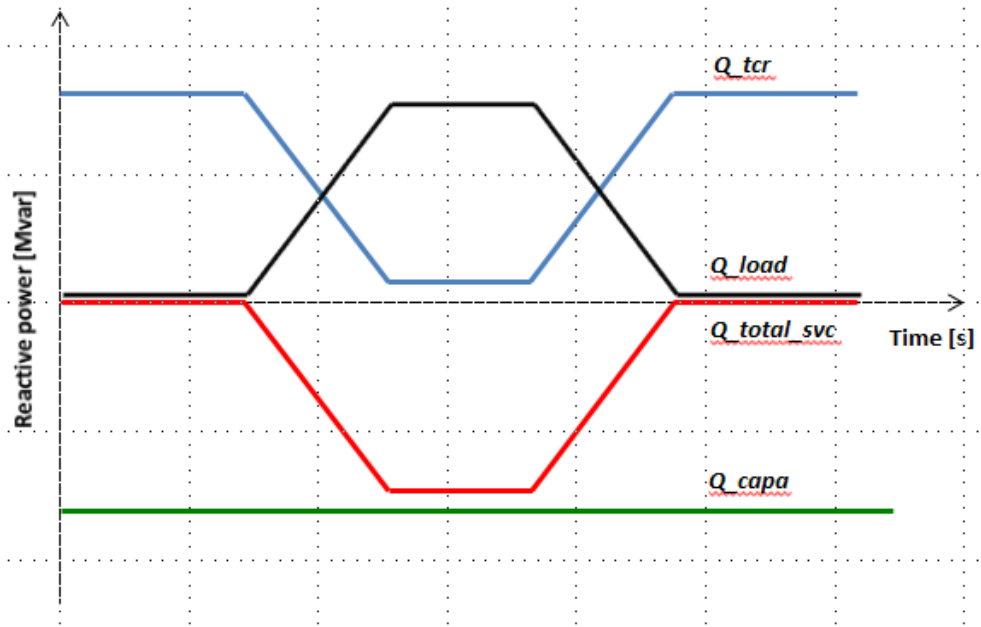

**Fig. 9:** Reactive power control of an SVC

It is a common misperception that SVCs store energy that is then used for the accelerator magnet load cycles. This is not the case! The reactive power circulation between the SVC (MVar generation) and the magnet power converters (MVar absorption) is a pure 50 Hz AC phenomenon, and no energy is stored or recovered between power cycles beyond one fundamental AC period of 20 ms.

An SVC cannot assure perfect voltage stabilization and perfect reactive power compensation at the same time. For most particle accelerators, power of the highest quality precedes the need for perfect reactive power compensation. The variable reactive power output of the SVC therefore needs to compensate not only the cycling reactive power of the load, but also correct the voltage variations caused by the cycling active power. Equation (6) defines the required reactive power function of the SVC.

$$Q_{\text{SVC}} = Q_{\text{load}} + \frac{P_{\text{load}}^2}{2S_{\text{cc}}} + kP_{\text{load}}, \tag{6}$$

where $k = R/X$ of the network.

### 6.1.1.1 *Harmonic filter design*

The total number of capacitors required for reactive power compensation are split into groups and then connected in series with air-core reactors to achieve the required harmonic tuning. The harmonic filters need to be designed taking into account the three major sources of harmonics:

- the 12-pulse thyristor power converter load (harmonic current source where $n = 11, 13, 23, 25$);
- the TCR (harmonic current source where $n = 5, 7, 11, 13, 17, 19$);
- the supplying external network, which is a harmonic voltage source typically where $n = 5$ and $7$ in steady-state, and $n = 3$ during transients.

The harmonic filters are tuned precisely to their corresponding harmonic frequency by adjusting the taps on the air-core filter reactors. Equation (7) calculates the resonance frequency of an L-C circuit:

$$f_{res} = \frac{1}{2\pi\sqrt{L_{filter}C_{filter}}} . \tag{7}$$

The total quantity of the capacitors will create a parallel resonance with the inductances of the supplying network. Equation (8) calculates the frequency of this parallel resonance:

$$f_{res} = f_0 \sqrt{\frac{S_{cc}}{Q_{SVC}}} . \tag{8}$$

For reliable SVC operation under all steady-state and transient conditions, the impedance of this parallel resonance needs to be controlled carefully. For large SVCs, it is typically in the range of 100–150 Hz, and hence would introduce instabilities during transient conditions. The installation of damped second and/or third harmonic filters can prevent or reduce amplification of these harmonics.

Based on the discussion above, an SVC for particle accelerators should hence consist of harmonic filters for $n = (2, 3,) 5, 7, 11, 13$ and high-pass. A typical topology is shown in Fig. 10. An impedance diagram for the harmonic filter topology shown in Fig. 10 is given in Fig. 11.

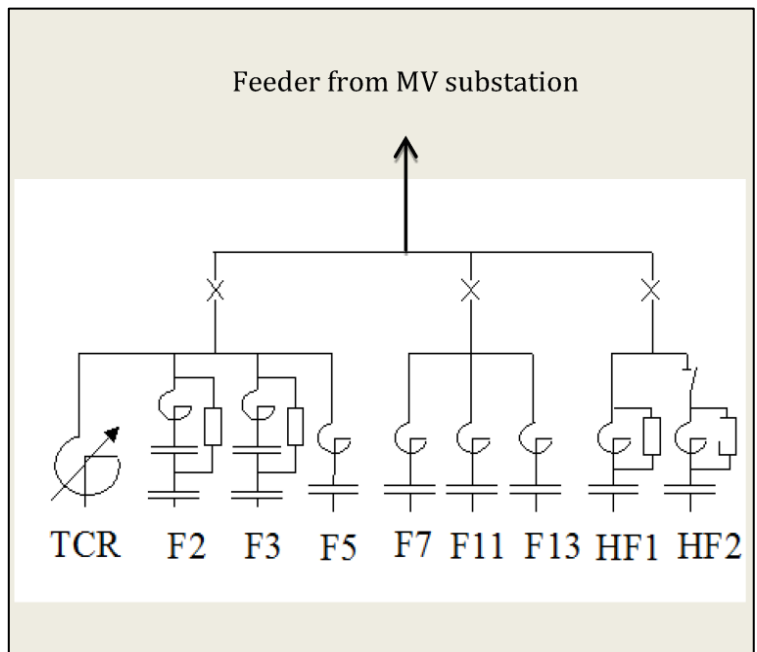

**Fig. 10:** Typical topology of an SVC for a particle accelerator

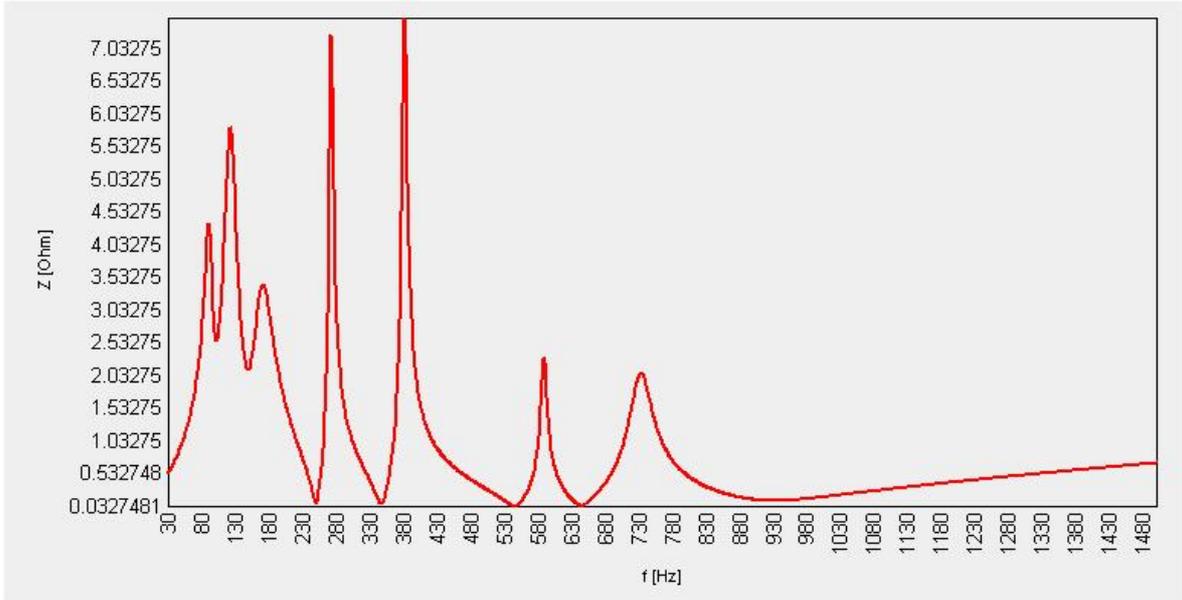

**Fig. 11:** Impedance diagram for the harmonic filter topology shown in Fig. 10

### 6.1.1.2  Control system

A typical SVC control system consists of a proportional integral (PI) AC voltage feedback controller and a direct compensation of disturbance to improve transient responses. As shown in Fig. 12, it includes three signals: load reactive power ($Q_{load}$), load reactive power differential ($dQ_{load}/dt$), and load active power ($P_{load}$), with all load variables being measured on the AC side of the power converter.

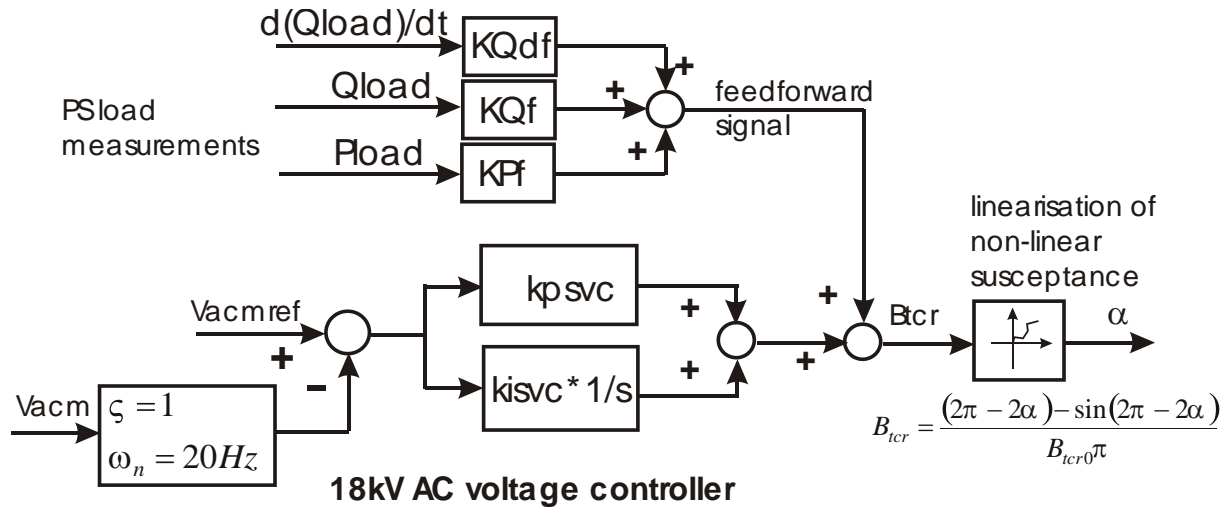

**Fig. 12:** Typical TCR control system

### 6.1.1.3  SVC performance

Typical SVC performance is discussed below, based on the example of the SPS accelerator. Figure 13 shows the cycles of active and reactive power for the load, the SVC, and the supplying network.

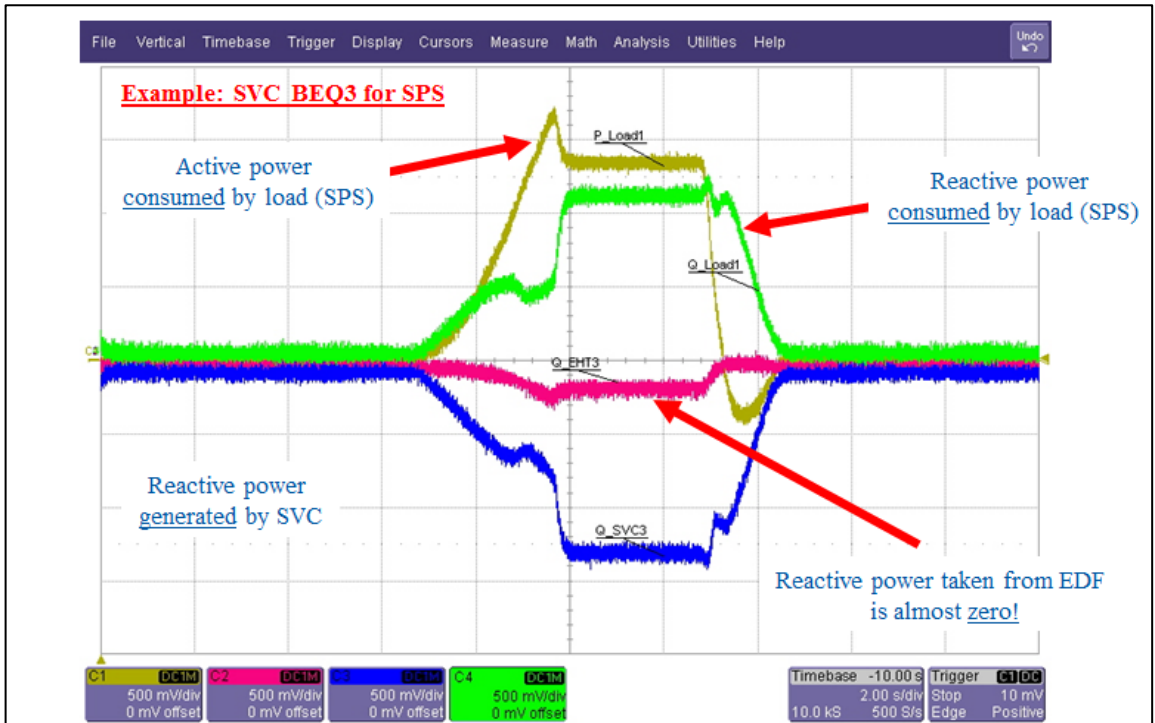

**Fig. 13:** Measured active and reactive load power cycles and reactive power output of the SVC (scale: 30 MW/MVar per division).

A well-tuned harmonic filter design as presented above would typically keep the THD of the bus voltage at the SVC connection point below 1%, during all phases of the cycling load pulse. Figure 14 shows the harmonic voltage distortion during the flat top of the cycle.

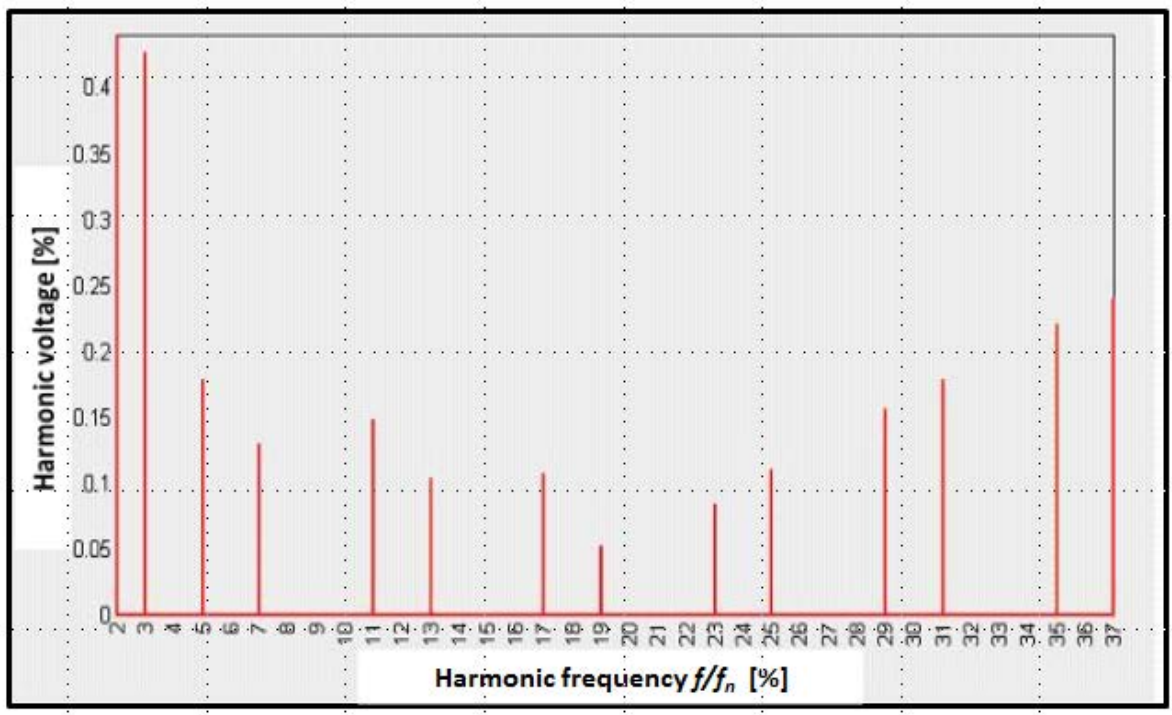

**Fig. 14:** Typical harmonic performance of an SVC with six harmonic filters, during flat top

The transient performance of the SVC response greatly depends on the response time of the control system, and also on the amplitude, shape, and rate of rise/fall of the active and reactive load cycles. Typically, the response time of an SVC is 50–100 ms. Figure 15 shows the measured voltage response of CERN's 18 kV network during SPS operation, with the SVCs in service. The voltage variations are within a limit of ±0.3% for most parts of the load cycle, and within ±0.75% during the transition points at the beginning and end of the load cycle.

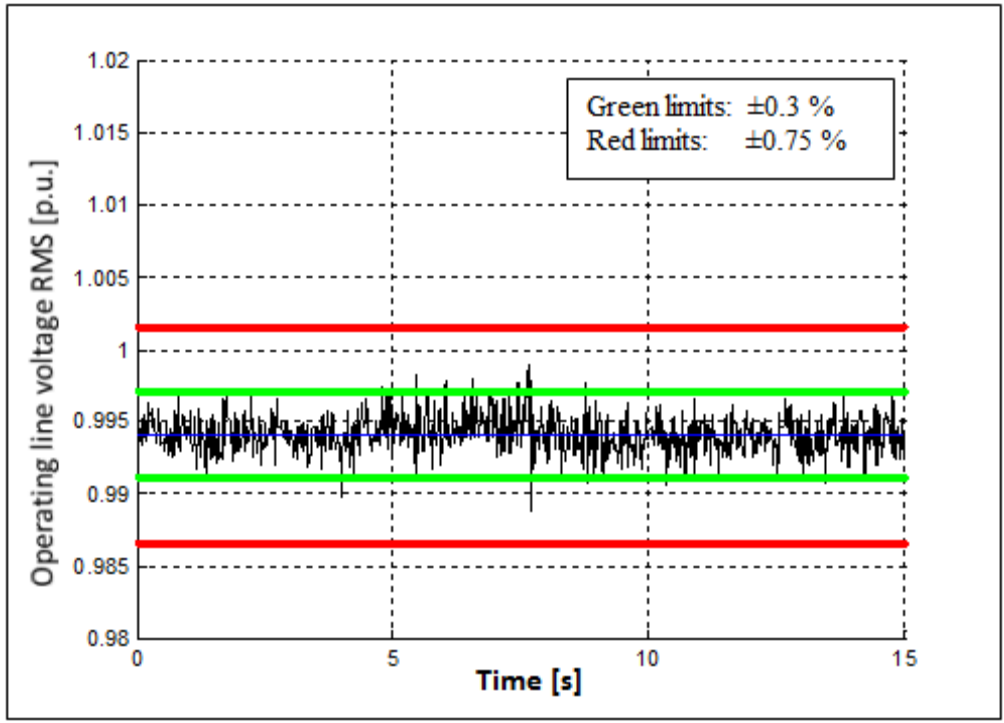

**Fig. 15:** Typical voltage response of an SVC compensating a large cycling load (e.g. SPS)

#### 6.1.2  *Static synchronous compensator without energy storage*

A static synchronous compensator (STATCOM) without energy storage covers functions similar to an SVC. Due to its switched-mode technology, its performance for voltage stabilization is superior when compared to an SVC. It is therefore an interesting technology for very fast cycling applications.

### 6.2  Systems with integrated energy storage

#### 6.2.1  *Power converters with integrated DC capacitor energy storage*

A novel approach for the decoupling of the load from the electrical network is the installation of power converters with integrated energy storage. The most prominent example of this technology is the new power supply for the PS (POPS) [14-16].

The POPS system consists of two active front end (AFE) rectifiers supplied from the 18 kV AC network, and six DC/DC converters. Associated with each DC/DC converter is one large DC capacitor bank for energy storage.

The six DC/DC converters are connected to the power system in two ways: two of them, being directly connected to the output of the two AFE's, are so-called 'chargers'. The four other DC/DC converters are 'floating'. The power part of these six DC/DC converters is identical; however both types of converters are controlled differently. Figure 16 shows the topology of the POPS system.

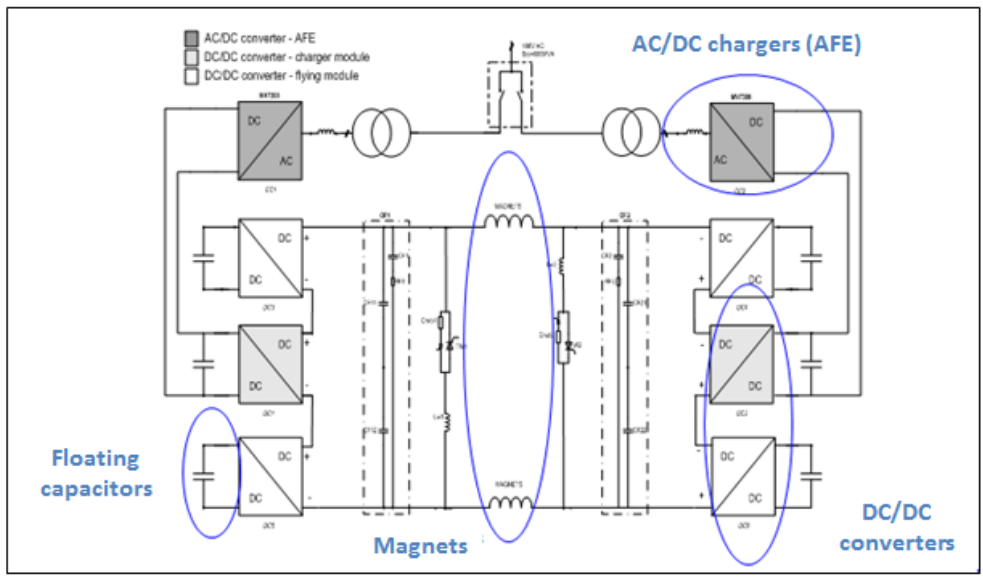

**Fig. 16**: POPS topology

On its DC output side POPS provides the current cycles required for the PS magnets (10 kV peak/6 kA peak). At the beginning of each cycle the DC storage capacitors are fully charged. During the load cycle they discharge, and then finally they recover a major part of the energy during the ramp-down of the accelerator magnet current towards the end of each cycle.

The two AFEs only cover the electrical losses of the power converters and accelerator magnets. To the electrical network, they represent an electrical load with a power factor of unity, and with only small variations in amplitude. As a consequence, the system does not generate any noticeable flicker in the electrical network.

The principal advantage of POPS is its modular design. Several degraded modes are integrated into its topology, permitting the filling of the LHC beam in the event of the failure of one power transformer, one AFE, or one DC/DC converter.

The POPS AFE and DC/DC converters are installed inside a dedicated power converter building. The cast-resin power transformers are installed outdoors, in metallic enclosures. The DC storage capacitor banks are installed in six sea shipping containers. A photograph of the POPS system is given in Fig. 17.

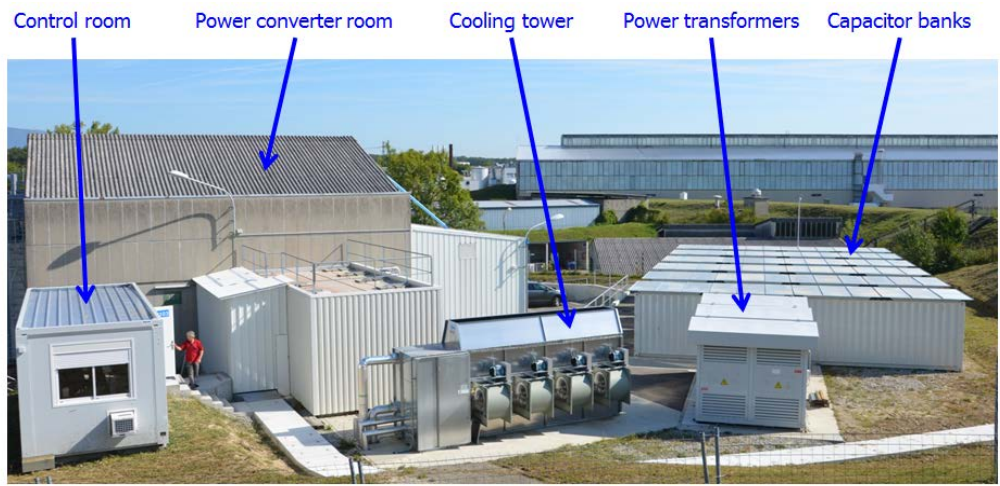

**Fig. 17**: POPS

Figure 18 shows the output voltage and current of POPS over one load cycle. One can distinguish three phases. During the ramp-up of the current (first phase), POPS operates at maximum output voltage. In the second phase, the current flat top, the voltage drops down to only cover the resistive losses of the magnets. Finally, during the third phase, the voltage is inversed to ramp down the current as quickly as possible. This inversion of voltage, while the current is still positive, corresponds to an inversion of power flow (from the magnets to the POPS system).

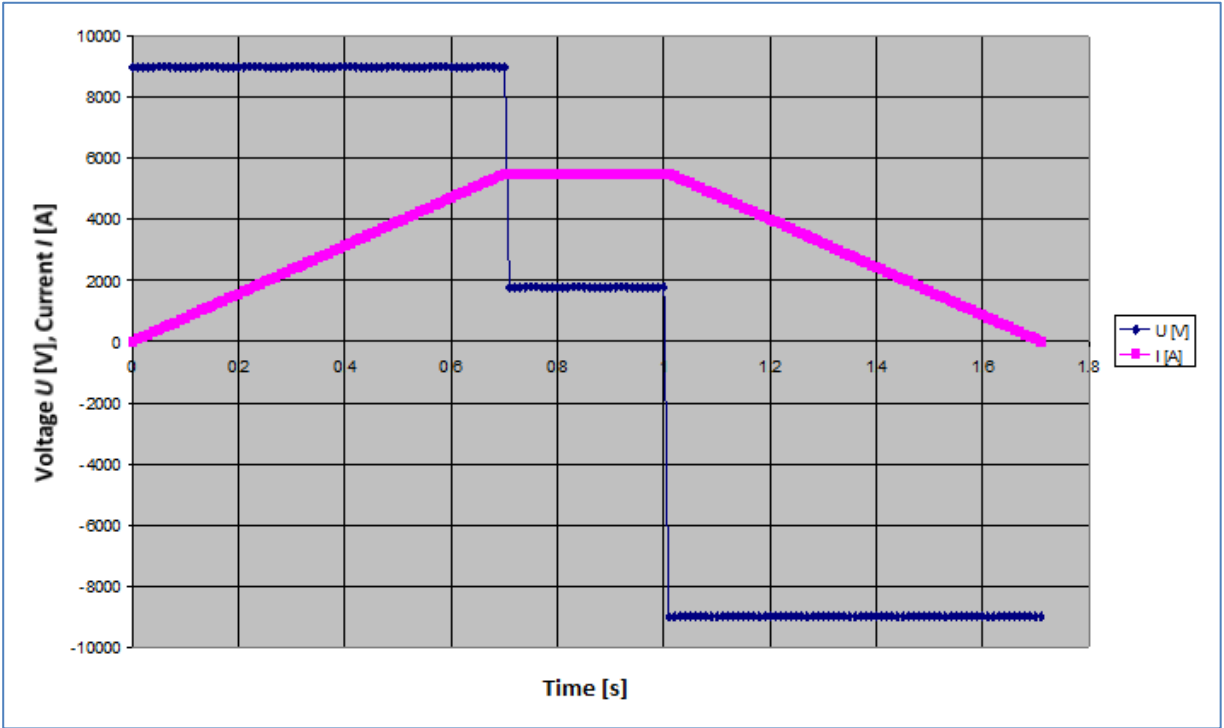

**Fig. 18**: POPS voltage and current output cycle

Figure 19 shows the active power cycle of the PS magnets, with a peak value of ±50 MW.

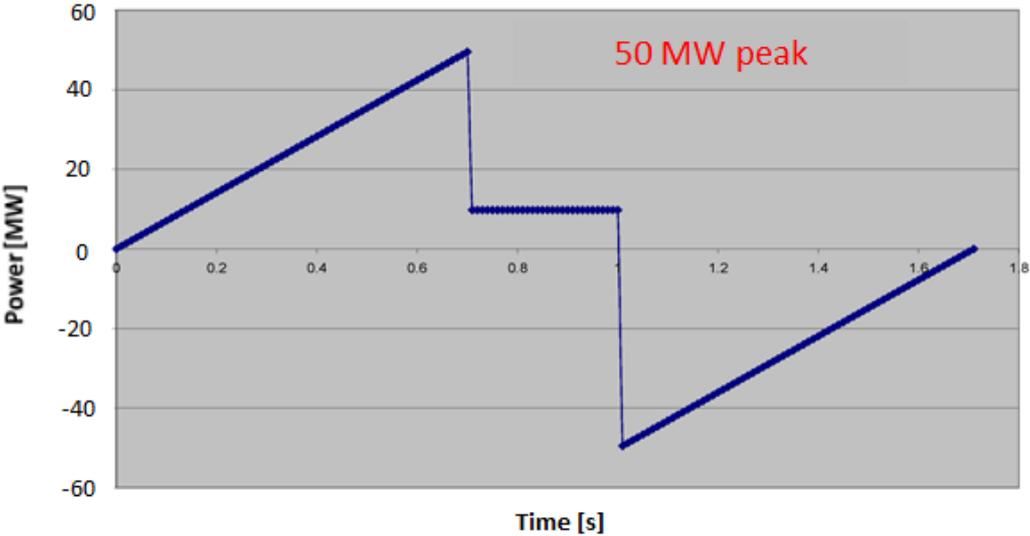

**Fig. 19**: Active power of the magnets

Figure 20 shows that the capacitors are fully charged at 5 kV DC at the beginning of the load cycle. They discharge during the power cycle and then recharge when the energy is fed back from the magnets to the POPS system.

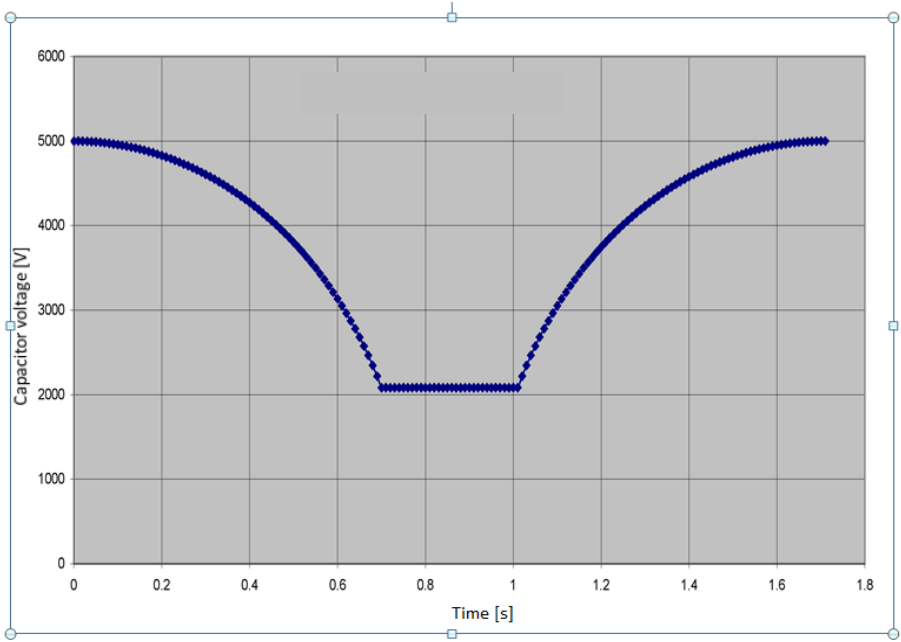

**Fig. 20:** Voltage profile of POPS DC storage capacitors during one load cycle

*6.2.1.1 Converter topology*

The AFE and DC/DC power converters consist of standardized industrial drives. To obtain the required current rating of 6 kA, three legs, each rated at 2 kA, are connected in parallel via coupling inductors. Each leg represents a three level neutral point clamped (NPC) branch.

Each AFE consists of one three-legged unit. Each DC/DC converter is an H-bridge with a three NPC leg unit on either side of the bridge for positive and negative polarity, as shown in Fig. 21. Each NPC leg consists of four insulated-gate bipolar transistors (IGBTs) and six diodes. The switching structure is supplied by two capacitors charged to $V_{dc}/2$. Each NPC leg allows the application of three voltage levels at the output ($V_{dc}/2$, 0, $-V_{dc}/2$).

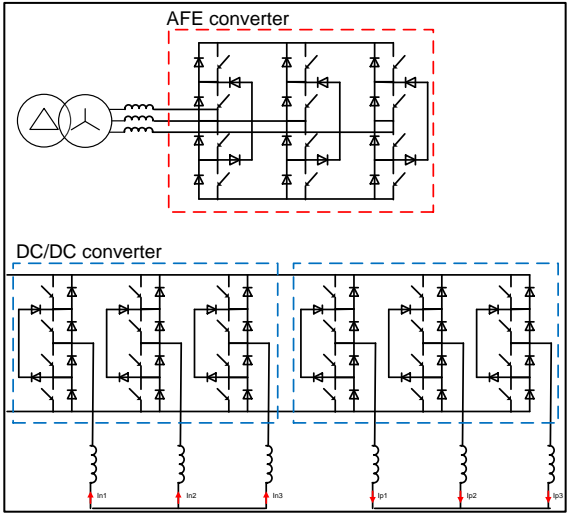

**Fig. 21**: AFE and DC/DC converter topologies

The different legs are interleaved, as are the converters, in order to increase the equivalent output ripple frequency, to reduce the size of the output filters, and to obtain better dynamics of the output voltage. For PS particle beam operation, the minimization of current ripple in the magnets is very critical. By using the principle of interleaving, an equivalent switching frequency of 4 kHz is obtained by using a basic switching frequency of 333 Hz for each individual leg. The interleaving between legs is obtained by shifting the modulation index of each leg by $2\pi/3$.

By using this topology, it was possible to assemble the entire POPS system by combining 14 standard industrial motor drives of 3.3 kV, 8 MW.

*6.2.1.2 Choice of semiconductors*

The PS represents a cycling load with a period of 1.2 s. POPS is specified for a lifetime of 15 years, corresponding to 100 million power cycles. The cycling character of the load is particularly important when choosing the type of semiconductor: the thermal time constant of the junction is in the order of 100 ms. During each load cycle longer than a few multiples of this time constant, the electrical connections to the junction will experience a certain ageing process due to thermal expansion. The prospective life time of the IGBT is a function of two conditions: the amplitude of junction temperature variation and the number of thermal cycles. Classical IGBTs, using very fine wire bonding for their electrical connections, are very susceptible to premature ageing due to these thermal cycles. Press-pack IGBTs, on the other hand, are a much better choice for cycling applications because they have no internally bonded wires. Here the internal electrical and thermal connections are done by applying very high compression forces by means of a clamp.

*6.2.1.3 Energy storage DC capacitor banks*

Each of the six capacitor banks has a rating of 0.25 F, 5 kV, and consists of 126 capacitor units in parallel. Each capacitor bank is installed inside a standard sea shipping container. The containers are air-conditioned to assure controlled ambient operating conditions for the capacitor banks. Figure 22 shows the inside of one of the containers.

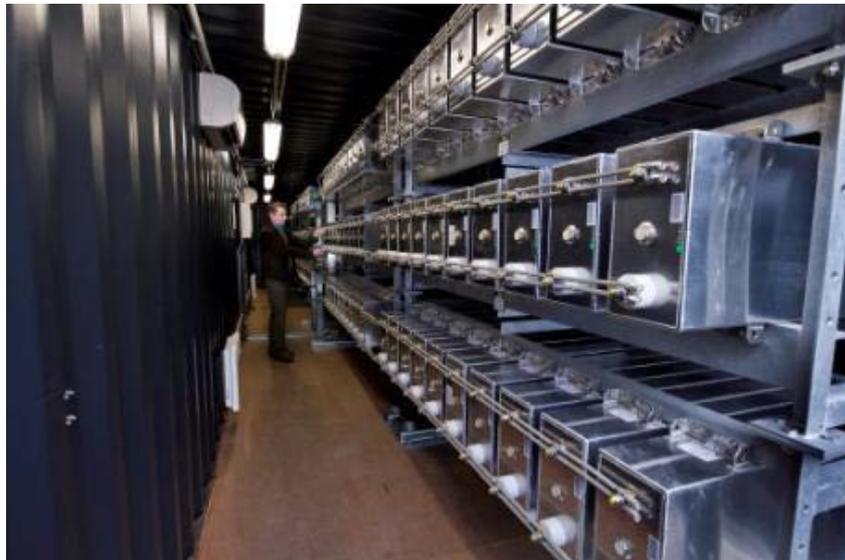

**Fig. 22**: Photo of one of the six DC storage capacitor banks inside its container

The DC capacitors are dry capacitors with polypropylene (PP) film. Small dielectric defects of the PP film are automatically repaired by 'self-healing', e.g. evaporation of the metalized electrode material around the dielectric defect. As an additional protection, in particular for major cases of capacitor failure, the capacitor bank is equipped with a fuse and an automatic discharge system.

### 6.2.1.4 Performance

Accuracy in B-field regulation is of particular importance during particle injection into and extraction from the PS. Figures 23 and 24 show the magnetic field in the dipole magnets during one load cycle.

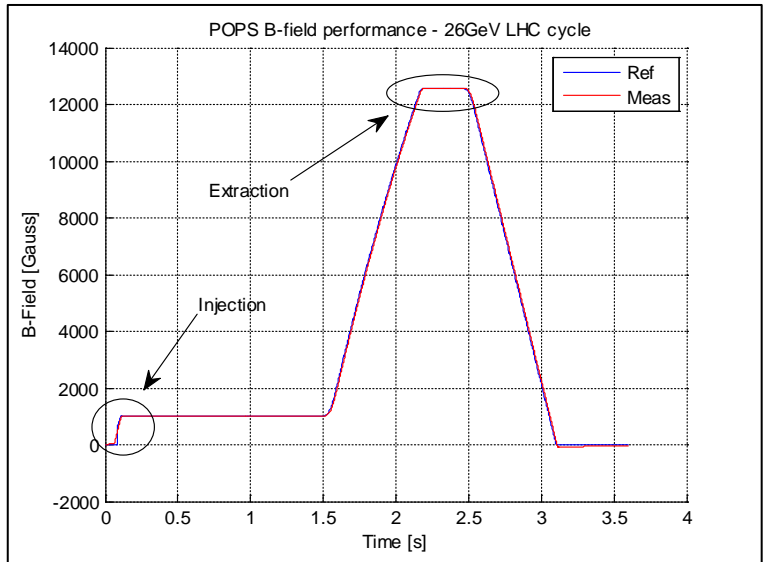

**Fig. 23**: POPS B-field regulation

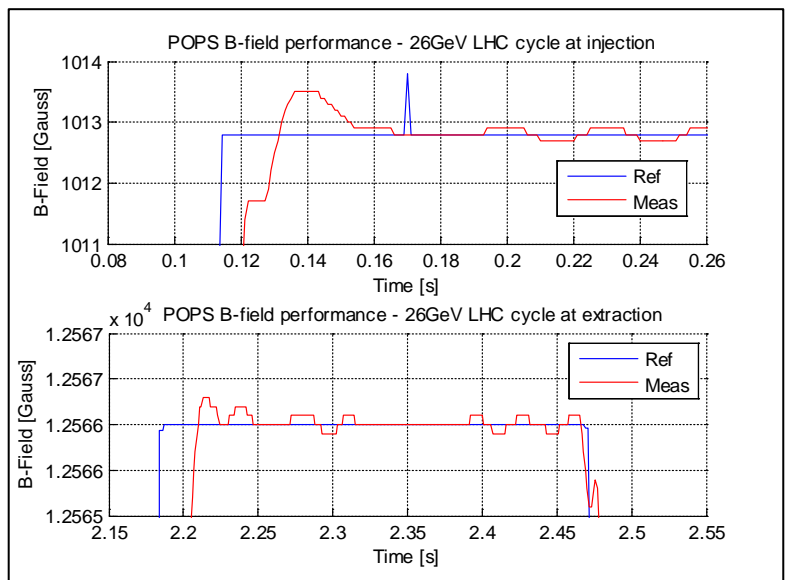

**Fig. 24**: Zoom of POPS B-field regulation

The digital RST controller reduces the error in magnetic field regulation of the magnets to 0.5 Gauss over an absolute value of 12000 Gauss, corresponding to an overall precision of $40 \times 10^{-6}$.

### 6.2.2 *Rotating machines for large cycling loads*

Another possibility for decoupling a cycling load from the electrical network, and to reduce flicker in the network, is the supply of a particle accelerator machine network via a rotating machine. This configuration is used for the AGS machine at BNL, and the PS rotating machine, which was in operation for 40 years before the commissioning of POPS.

A motor drives a large synchronous generator that supplies the cycling power converter load. Due to the kinetic energy stored in the rotation of the motor-generator set, the power drawn from the electrical network by the motor follows the accelerator cycles with relatively small power variations.

These systems have reliably operated for many years; however they always represent a considerable risk as a single point of failure. A major failure of the rotor could easily stop the operation of the machine for at least six months. Figure 25 shows a single-line diagram of the PS rotating machine at CERN, while Fig. 26 gives one of its typical load cycles.

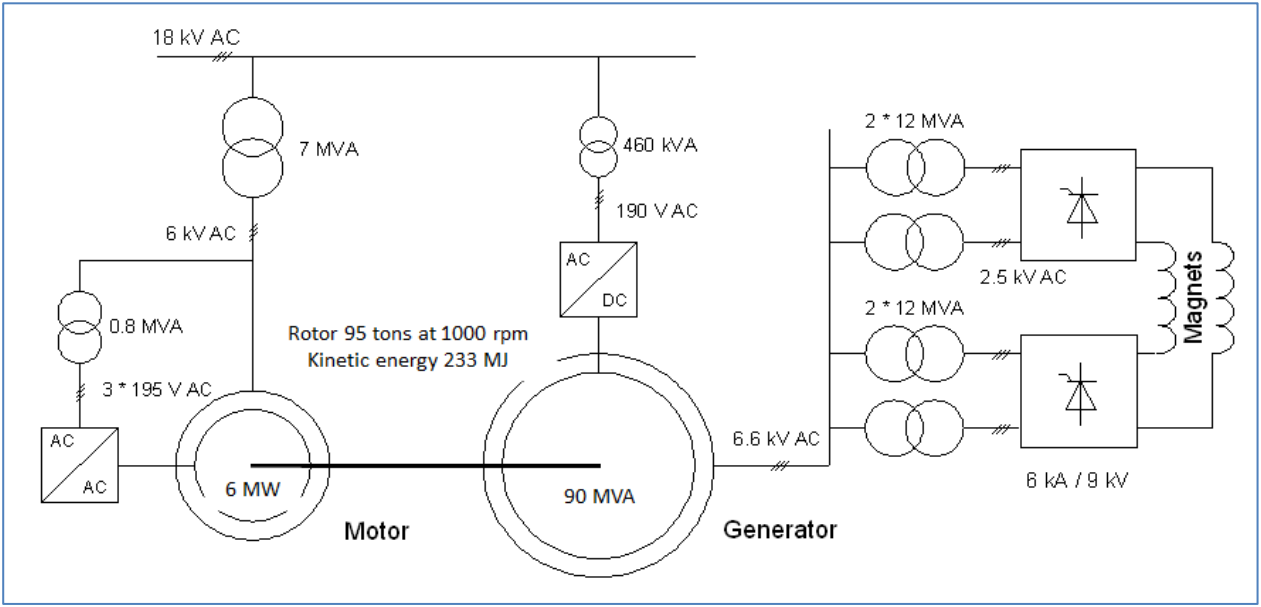

**Fig. 25**: Single-line diagram of the PS rotating machine at CERN

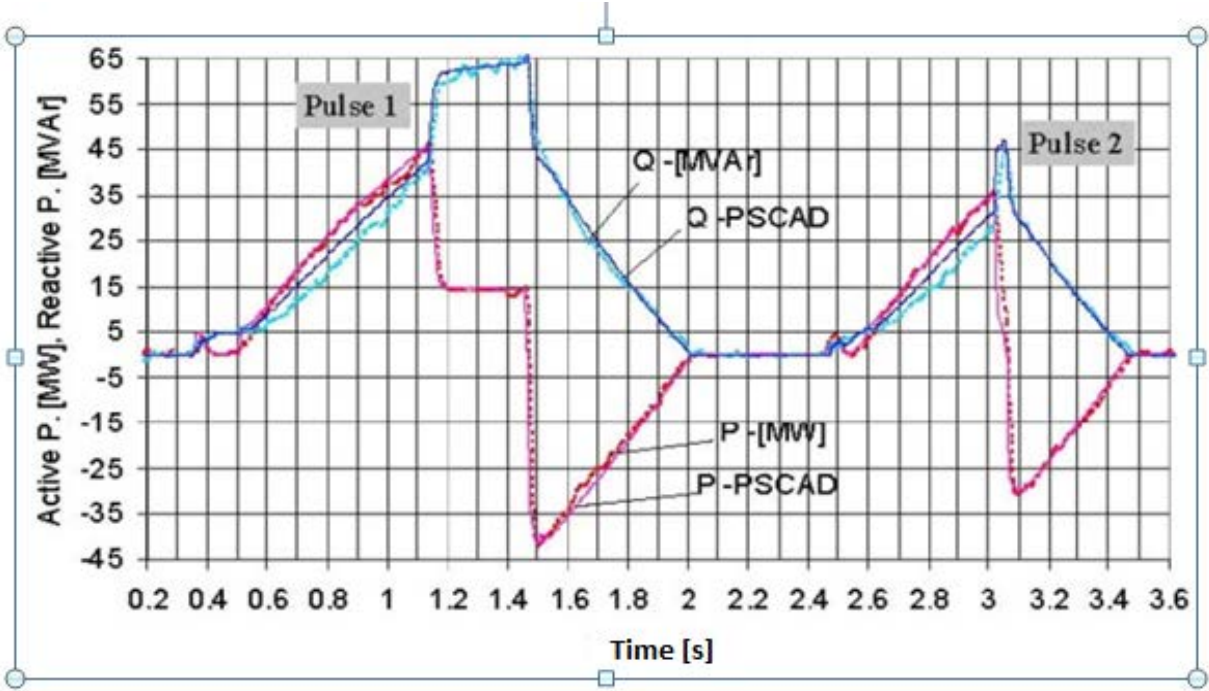

**Fig. 26**: Typical active and reactive power cycle of the PS main power converter

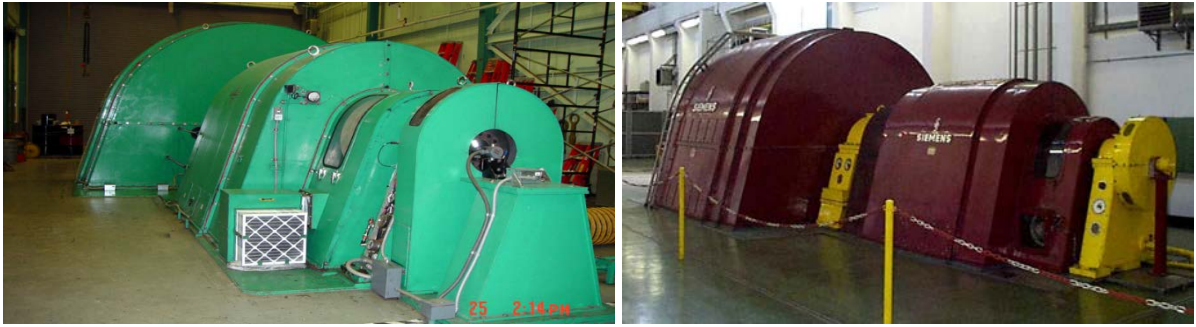

(a) (b)
**Fig. 27:** (a) AGS machine at BNL; (b) PS machine at CERN

*6.2.3 Flywheel system to compensate external network disturbances*

Physics laboratories often suffer from electrical network disturbances, and many of these events result in particle accelerator stoppages. Due to its geographical location in Grenoble, surrounded by the French Western Alps, ESRF is particularly exposed to disturbances caused by thunderstorms, which are most frequent during the summer months. Figure 28 shows the seasonal distribution of voltage disturbances [17].

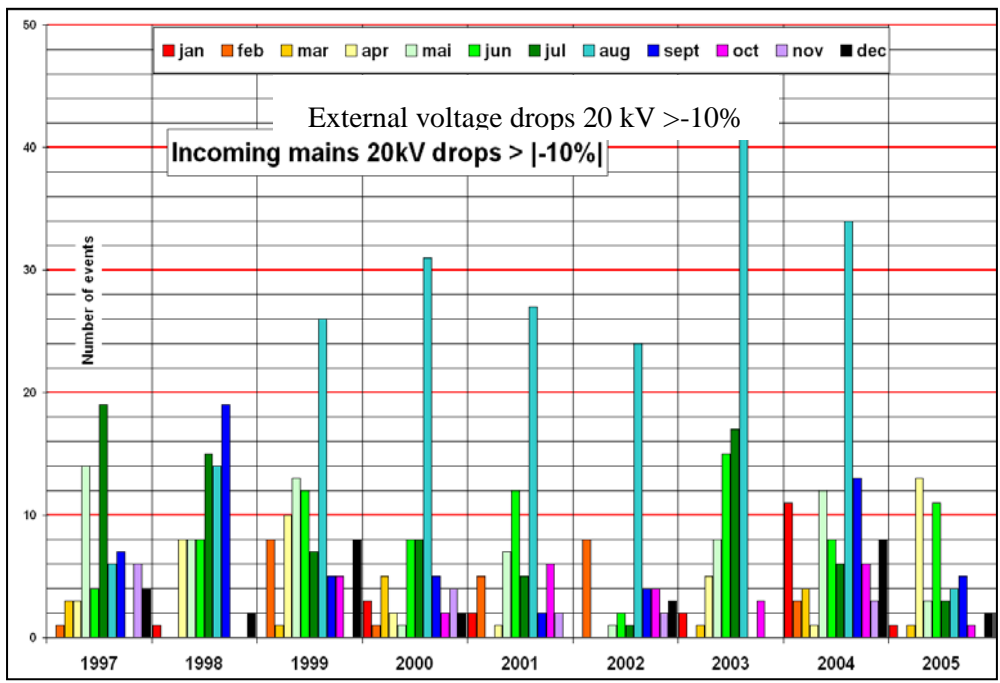

**Fig. 28**: Recorded frequency of voltage disturbances

In order to increase the mean time between failure (MTBF), ESRF decoupled the 20 kV accelerator network from the external supply by installing several sets of rotating machines, each consisting of an alternator and rotating mass (accumulator).

For smaller voltage disturbances (conditioning zone), the rotating machines just provide the difference to re-establish a fully balanced system with 100% of the voltage amplitude. In the event of disturbances with a larger amplitude or longer duration (disconnection area), the accelerator network is automatically disconnected from the mains and fully supplied by the rotating machines. The total stored kinetic energy of all machines is 100 MJ, and the system can compensate for 100% of missing power for 12 s. Figure 29 shows the limits of the conditioning and disconnection areas [17].

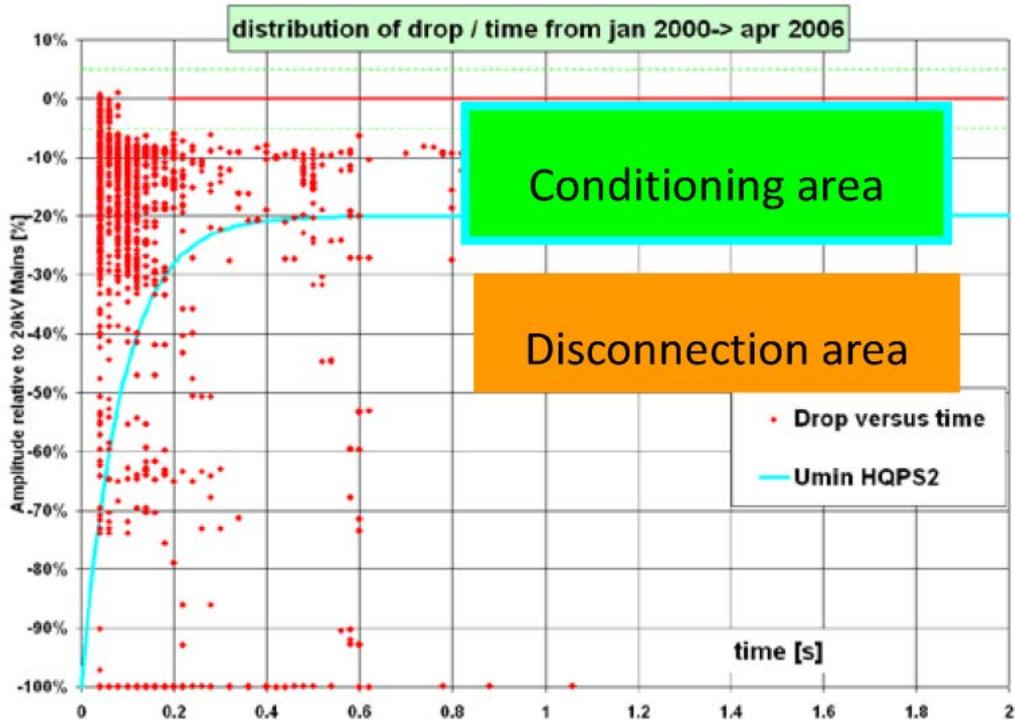

**Fig. 29**: Conditioning and disconnection area

Due to the combination of several machines, the system has a built-in redundancy and will also continue to operate in the event of the unavailability of one or two units. Figure 30 shows a general diagram of the system [17].

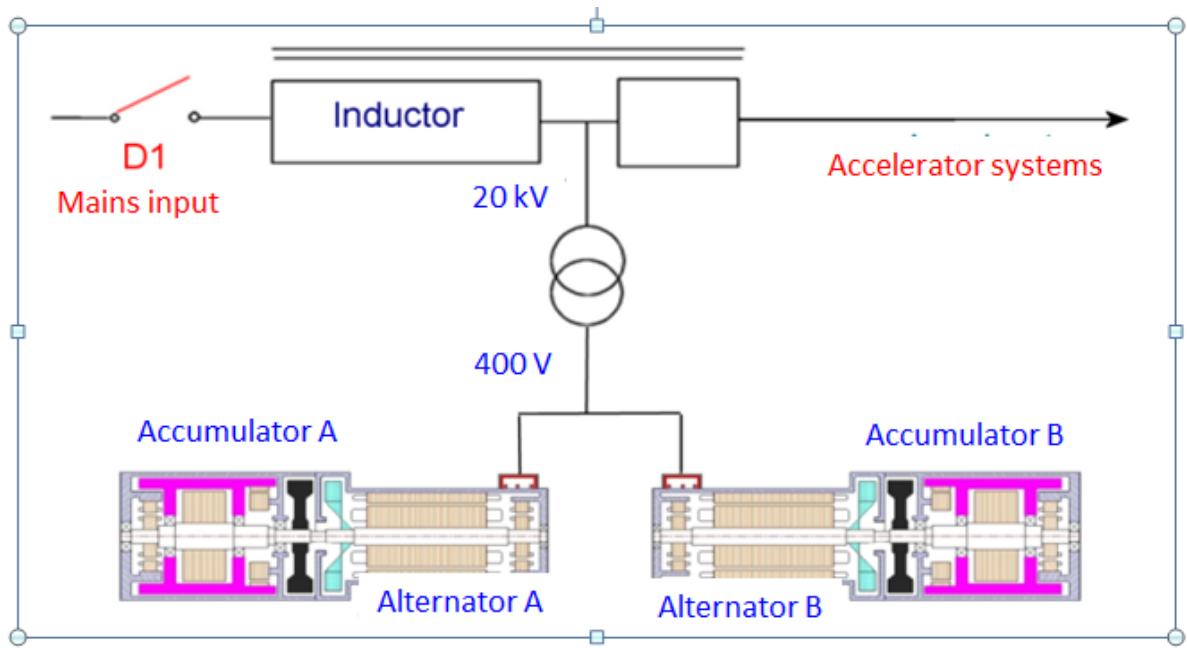

**Fig. 30**: Electric diagram of the network conditioning system

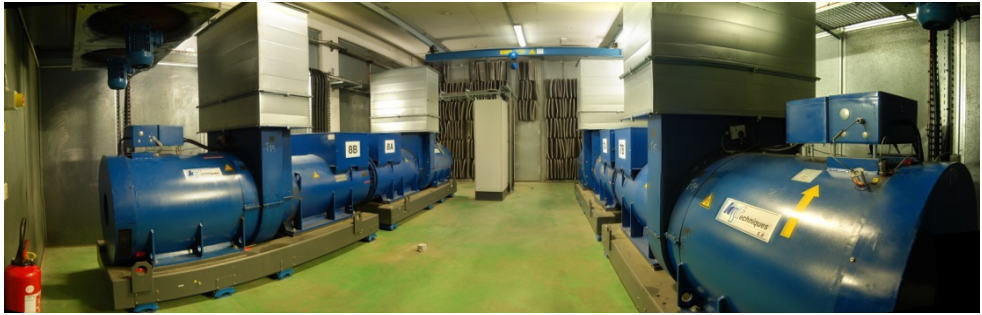
**Fig. 31**: Two twin rotablocs (four accumulators and four alternators in one cell)

*6.2.4    Static synchronous compensator with energy storage*

The STATCOM controls a variable output of active and reactive power, and hence completely decouples the cycling load from the electrical network. The energy could be stored in HV DC capacitors or in a superconducting magnet coil (SMES). The STATCOM technology is particularly suitable for connecting large cycling power converter loads to relatively weak electrical networks. A STATCOM could potentially also be used to compensate for transient voltage disturbances in an accelerator network.

This technology has not yet been applied for a particle accelerator network, however technical studies have commenced to evaluate the potential for this technology at CERN [18, 19].

# 7     Summary and recommendations

Particle accelerators require the control of the magnet current with the highest precision, sometimes to a few parts per million. To make excellent physics, excellent power quality is required! The most important aspects and recommendation of this document are summarized below.

- Voltage disturbances are part of the normal operation of electrical networks. They are frequent, and equipment must possess a certain immunity to correctly function in this environment.

- International standards such as IEC 61000 define major aspects of electromagnetic compatibility and power quality. However, in some cases they do not sufficiently cover the specific needs of your physics laboratory. In those cases *you* need to define the principal immunity levels for your electrical equipment!

- Identify the type and frequency of network disturbances that cause most of your accelerator stops, and establish detailed statistics to better understand them.

- All groups installing and operating electrical equipment need to be involved in power quality considerations, right from the beginning of a project.

- Strictly separate (cycling) power converter loads from general services loads; supply these networks from separate power transformers.

- Minimize network impedances (inductances!) to reduce voltage variations and harmonic distortion in your networks.

- When choosing a power converter topology, aim to minimize the amplitude of *cycling* reactive and active power to minimize the generation of flicker.